\documentclass{amsart}
\usepackage{amsmath}

\newtheorem{lemma}{\bf Lemma}[section]
\newtheorem{theorem}{\bf Theorem}[section]
\newtheorem{corollary}{\bf Corollary}[section]
\newtheorem{proposition}{\bf Proposition}[section]
\newtheorem{example}{\bf Example}[section]
\newtheorem{definition}{\bf Definition}[section]
\newtheorem{remark}{\bf Remark}[section]

\def\La{\Leftarrow}
\def\o{\overline}
\def\laa{\leftarrow}
\def\Ra{\Rightarrow}

\title[On finite  functions with non-trivial arity gap]
 {On finite  functions with non-trivial arity gap}
\begin{document}
\author{{Slavcho Shtrakov }  \and {J\"org  Koppitz}}\thanks{The research  was supported by the DFG project KO 1446/3-1}

\address{Department of Computer Science \\South-West University,  2700 Blagoevgrad, Bulgaria,} \email{ shtrakov@swu.bg}
\urladdr{http://home.swu.bg/shtrakov}
\address{Institute of Mathematics, University of
Potsdam, 14415 Potsdam,}  \email{ koppitz@rz.uni-potsdam.de}
\date{}
\keywords{essential variable, identification minor, essential arity
gap.}
  \subjclass[2000]{Primary:  03G25; Secondary: 05E05\\
~~~~\ \ \ \ ~~~~~~~~~ \it ACM-Computing Classification System (1998) : G.2.0}

\begin{abstract}
Given an $n$-ary
  $k-$valued function  $f$, $gap(f)$ denotes the minimal number of essential variables in $f$ which become fictive when identifying any two distinct essential variables in $f$.

We particularly solve  a problem  concerning the explicit determination of  $n$-ary
  $k-$valued functions $f$ with $2\leq gap(f)\leq n\leq k$. Our methods yield new combinatorial results about the number of such functions.
\end{abstract}

\maketitle

\section{Introduction}
~~~\\ \\

Given a function $f$, the essential variables in $f$ are defined as variables which occur in $f$ and weigh with the values of that function.
  The number of   essential  variables is an important measure of complexity for discrete functions. Some deep results characterizing such variables and sets of essential variables are known
(\cite{bre,ch51,kov,lup,sal}). Similar problems about terms in universal algebra are discussed  in  \cite{den,sh51,sh52} and about
tree automata    in \cite{s1}. So, any opportunity to  reduce  the number of essential variables in discrete functions is an important procedure
 in  theoretical and applied computer science and
modeling. There are two ways to decrease this number - by replacing some variables in function with constants or with other variables (i.e. with identification of variables).

 M. Couceiro and E. Lehtonen classify finite valued functions on a finite set $A$ in terms of their arity gap (\cite{mig1}). The aim of the present paper is the representation  and description of such functions. We shall use  the well known fact that each $n$-ary function $f: A^n\to A$ can be represented as sums of conjunctions.

In  Section \ref{sec2} we introduce the basic definitions and some
preliminary results concerning $k-$valued functions.

In Section \ref{sec3} we study the essential arity gap of functions.
Here  a complete description
of the functions in $k$-valued logic depending essentially on all of its $n$ variables
whose  essential arity gap is equal to $p$
with $2< p\leq n\leq k$ is obtained.

In  Section \ref{sec6} we consider the class $G_{2,k}^n$  of $n$-ary $k$-valued functions which have essential arity gap equal to 2 when $k>2$. This class is presented as union of two subclasses which are investigated.

In  Section \ref{sec5} we discuss a special class of the ternary $k$-valued functions which have essential arity gap equal to 2. The number of  functions in this class is found.

\section{Preliminaries}~\label{sec2}
~~~~\\ \\

 Let $k>2$ be a natural number.
 Denote by $K=\{0,1,\ldots,k-1\}$ the set (ring) of
 remainders modulo $k$. A  function
(operation) on $K$  is a mapping $f: K^n\to K$ where $n$ is a  natural
number, called {\it the arity} of $f$. The set of the all such
functions is denoted by $P_k^n$. Operations from $P_2^n$ are called
{\it Boolean functions}.

Let  $X_n=\{x_1,\ldots,x_n\}$  be the set of $n$ variables and  $f\in P_k^n$ be a $k$-valued function.
\begin{definition}\label{d1.2}
A variable $x_i$ is called {\it essential} in $f$, or $f$ {\it
essentially depends} on $x_i$, if there exist values
$a_1,\ldots,a_n,b\in K$, such that
$$
   f(a_1,\ldots,a_{i-1},a_{i},a_{i+1},\ldots,a_n)\neq
   f(a_1,\ldots,a_{i-1},b,a_{i+1},\ldots,a_n).
$$
\end{definition}
The set of all essential variables in a function $f$ is denoted by
$Ess(f)$ and the number of its essential variables  is denoted by
$ess(f)=|Ess(f)|$. The variables from $X_n$ which
are not essential in
 $f\in P_k^n$ are called {\it fictive} and the set of fictive
 variables in $f$ is denoted by $Fic(f)$.

Let $x_i$ and  $x_j$ be two distinct essential variables in $f$. We say that the
function $g$ is obtained from $f\in P_{k}^{n}$ by {\it the
identification of the variable $x_i$ with $x_j$}, if
$$
   g(a_1,\ldots,a_{i-1},a_i,a_{i+1},\ldots,a_n)=f(a_1,\ldots,a_{i-1},a_j,a_{i+1},\ldots,a_n),
$$
for all $(a_1,\ldots,a_n)\in K^n$.

Briefly, when  $g$ is obtained from $f, $ by identification of the
variable $x_i$ with $x_j$, we will write $g=f_{i\laa j}$ and $g$ is
called {\it the identification minor of $f$} and  $Min(f)$ denotes the set of all
identification minors of $f$.

We shall allow  formation of identification
minors when $x_i$ or $x_j$ are not essential in $f$, also. Such minors of $f$ are called {\it trivial} and they do not belong to $Min(f)$.  For instance, if
$x_i$ does not occur in $f$, then  $f_{i\laa j}:=f$.

\begin{remark}\label{rem2.1}
Let  $i$ and $j$ be two natural numbers with $1\leq j,i\leq n$, $i\neq j$. Then we have:

$(i)$   $ess(f_{i\laa j})\leq ess(f)$, because
$x_i\notin Ess(f_{i\laa j})$, even though it might be essential in
$f$.

$(ii)$ If $x_j$ is an essential variable in $f$ then $Ess(f_{i\laa j})\subseteq Ess(f)$.
\end{remark}
\begin{definition}\label{d1.1}
Let  $f\in P_k^n$ be an $n$-ary $k$-valued function. Then   {\it the essential arity gap} (briefly
{\it arity gap} or {\it  gap}) of $f$ is defined as follows
$$gap(f):=ess(f)-\max_{g\in Min(f)}ess(g).$$
\end{definition}

We say that the function $f$ has non-trivial arity gap if $gap(f)\geq 2$.

We let $G_{p,k}^m$ denote the set of all functions in $P_k^n$
which essentially  depend on $m$ variables whose arity gap is equal to 
$p$ i.e. $G_{p,k}^m=\{f\in P_k^n\ |\ ess(f)=m\ \&\ gap(f)=p\}$, with
$m\leq n$.

The set of all $n$-ary $k-$valued functions which essentially depend
 on $m$  variables is denoted by $P_{m,k}^n$, i.e. $P_{m,k}^n=\{f\in P_k^n\ |\ ess(f)= m\}$, $m\leq n$.

An upper bound of $gap(f)$ for Boolean functions is found by
K.Chimev, A. Salomaa and  O. Lupanov  \cite{ch3,sal,lup}. It is
proved that $gap(f)\leq 2$, when  $f\in P_2^n,\ n\geq 2$.

This result is generalized for arbitrary finite valued functions in
\cite{mig0}. It is shown that $gap(f)\leq k$ for all $f\in P_k^n$.

In \cite{mig0} the  Boolean functions whose arity gap is equal to  2 are described.
 In \cite{s2} the class  $G_{2,2}^n$ is  investigated, also.   Several combinatorial results
concerning the number of the functions in this class are obtained.

The case $2\leq p\leq k<n$  is fully described in \cite{ros} where it is proved
that $gap(f)\leq 2$ and if $f\in G_{2,k}^n$ then $f$ is a totally symmetric function.

So, in the present paper  we  shall pay attention to  the case
$2<  k$ and $n\leq k$.

We show that if $f\in G_{p,k}^n$, $2<p\leq n\leq k$ then $f=h\oplus g$ where $ess(h)=n-p$ and  $g\in G_{n,k}^n$ which arises in this way, solving the following problem:\\
{\it For each $1\leq p \leq k$, determine explicitly the functions $f\in  P_k^n $
whose arity gap is equal to  $p$, $p\geq 2$} (\cite{mig0}, p. 6, Problem 1).

Let $m\in N$, $0\leq m\leq k^n-1$ be an integer. It is well known
that for every $k,n\in N,\ k\geq 2$ there is an unique finite
sequence $(\alpha_1,\ldots,\alpha_n)\in K^n$ such that
\begin{equation}\label{eq2.1}
  m=\alpha_1 k^{n-1}+ \alpha_2 k^{n-2}+\ldots+\alpha_n.
\end{equation}
The equation (\ref{eq2.1}) is known as the representation of $m$ in
$k-$ary positional numerical system. One briefly  writes   $
m=\o{\alpha_1\alpha_2\ldots\alpha_n}$ instead (\ref{eq2.1}).

Given a variable $x$ and $\alpha\in K$, $x^\alpha$ is an important
 function defined by:
$$
  x^\alpha=\left\{\begin{array}{ccc}
             1 \  &\  if \  &\  x=\alpha \\
             0 & if & x\neq\alpha.
           \end{array}
           \right.
$$

 In this paper we shall use {\it sums of
conjunctions (SC)} for representation of functions in $P_k^n$. This is the most natural representation of the functions in finite algebras. It is based on so called operation tables of the functions.
\begin{theorem}\label{t2.1}
Each function $f\in P_k^n$ can be uniquely represented in SC-form as
follows
\begin{equation}\label{eq2.2}
    f=a_0.x_1^{0}\ldots x_n^{0}\oplus\ldots\oplus
     a_{m}.x_1^{\alpha_1}\ldots
     x_n^{\alpha_n}\oplus\ldots\oplus a_{k^n-1}.x_1^{k-1}\ldots
     x_n^{k-1}
\end{equation}
with  $m={\o{\alpha_1\ldots\alpha_n}}$, $a_{m}\in K$, where $"\oplus"$ and
$"."$ are the operations addition and multiplication modulo $k$ in
the ring $K$.
\end{theorem}

A fact we shall use repeatedly is this:
A variable $x_i$ is fictive (inessential) in the function  $f\in P_k^n$,  if and
only if
$$
    f(x_1,\ldots,x_n)=
    x_i^{0}.f_1\ \oplus\ x_i^{1}.f_2\ldots
     \oplus x_i^{k-1}.f_k,
$$
with $f_1=f_2=\ldots=f_k$ and $x_i\notin Ess(f_j)$, where $f_j$ are
$n-1$-ary $k$-valued functions with set of variables
$\{x_1,\ldots,x_{i-1},x_{i+1},\ldots,x_n\}$ for  $j=1,2,\ldots,k$.

Consequently,
if $f,g\in P_k^n$, $x_i\in Fic(f)$ and $x_i\in Fic(g)$, then $x_i\in Fic(f\oplus g)$.

Let us note that if   $f_{i\laa j}=g_{i\laa j}$ and
 $\alpha_i=\alpha_j$ then $f(\alpha_1,\ldots,\alpha_j,\ldots,\alpha_i,\ldots,
\alpha_n)=g(\alpha_1,\ldots,\alpha_j,\ldots,\alpha_i,\ldots,
\alpha_n)$.

\begin{lemma}\label{l2.2}
Let $f$ and $g $ be two  $k$-valued functions, depending essentially on
$n,\ n\geq k+1$ variables. If   $f_{i\laa j}=g_{i\laa j}$ for all $i,j$ with
$ 1\leq j , i\leq n$ and $i\neq j$, then $f=g$.
\end{lemma}
\begin{proof} Let $(\alpha_1,\alpha_2,\ldots,
\alpha_n)\in K^n$ be an arbitrary $n$-tuple of integers from $K$. Since $n\geq k+1$ it follows that there exist two natural numbers $i,\ j$ with $1\leq j<i\leq
 n$ and $\alpha_i=\alpha_j$. Then $f_{i\laa j}=g_{i\laa j}$  implies
  $$f(\alpha_1,\ldots,\alpha_j,\ldots,\alpha_i,\ldots,
\alpha_n)=g(\alpha_1,\ldots,\alpha_j,\ldots,\alpha_i,\ldots,
\alpha_n).$$ Consequently
  $f=g$.
 \hfill ~~~ \end{proof}
\begin{example}\label{e2.1}
Let us consider the functions $f=x_1^0x_2^0x_3^0\oplus
x_1^1x_2^0x_3^2$ and $g=x_1^0x_2^0x_3^0\oplus x_1^0x_2^1x_3^2$ from
$P_3^3$. It is not too hard to show that $f_{i\laa j}=g_{i\laa
j}=x_j^0x_m^0$, where $m\in \{1,2,3\}\setminus\{i,j\}$ for all $i,j$ with $
1\leq j , i\leq 3$ and $i\neq j$. On the other hand we have $f\neq g$. This example
shows that $n\geq k+1$ is an essential  condition in Lemma \ref{l2.2}.
\end{example}
\begin{lemma}~\label{l2.3}
 If $x_i\notin Ess(f)$ and $f\in P_k^n$ then
$f=f_{i\laa j}$, for all $j\in \{1\ldots n\}$, $i\neq j$.
 \end{lemma}

\begin{proof}\ Without loss of generality assume that $j=1$ and $i=2$. Then we have
$$
f(\alpha_1,\alpha_2,\alpha_3,\ldots,\alpha_n)=f(\alpha_1,\beta_2,\alpha_3,\ldots,\alpha_n),
$$
for all $\alpha_1,\alpha_2,\alpha_3,\ldots,\alpha_n,\beta_2\in K$.
Let
$\gamma_1,\gamma_2,\gamma_3,\ldots,\gamma_n\in K$  be arbitrary $n$ integers from $K$.
Then we have
$$ f(\gamma_1,\gamma_1,\gamma_3,\ldots,\gamma_n)=f(\gamma_1,\gamma_2,\gamma_3,\ldots,\gamma_n).
$$
Consequently $f=f_{2\laa1}$.
 \hfill ~~~ \end{proof}

Lemma \ref{l2.3} implies that
if $x_i\notin Ess(f)$ then
$Ess(f)=Ess(f_{i\laa j})$, for all $j\in \{1\ldots n\}$ with $i\neq j$.

\section{Essential arity gap of $k$-valued
functions}~~\label{sec3}%, depending essentially on $k$-variables}
~~~~
\\ \\
 We are going to study the $n$-ary  $k$-valued functions whose arity gap is equal to  $n$. The set of all strings over $K$ with length $m$, $m\geq 1$ will be denoted by $K^m$.

Given two natural numbers $k,n\geq 2$, $Eq_k^n$ denotes the set of
all strings over $K=\{0,1,\ldots,k-1\}$ with length $n$ which have
at least two equal letters i.e.
$$Eq_k^n:=\{\alpha_1\ldots\alpha_n\in K^n\  |\ \ \alpha_i=\alpha_j,
\mbox{ for some } i,j\leq n,\ i\neq j\}.$$

\begin{lemma}~\label{c33.1}
If $f\in G_{n,k}^n$ and $2\leq n\leq k$, then
$f(\alpha_1,\ldots,\alpha_n)=f(0,\ldots,0)$ for all  $\alpha_1\ldots\alpha_n\in Eq_k^n$.
\end{lemma}

\begin{proof} Let $\alpha_1\alpha_2\ldots\alpha_n$ be an arbitrary string from $ Eq_k^n$.
Without loss of generality let us assume that $\alpha _{1}=\alpha _{2}$.
Since $f\in G_{n,k}^{n}$ it follows that the function $f_{2\leftarrow 1}=
f(x_1,x_1,x_3,\ldots,x_n)$ does not essentially depend on any
of its variables $x_{1},x_{3},\ldots, x_{n}$. So, we have $f(\alpha _{1},\alpha _{2},\alpha _{3},\ldots
,\alpha _{n})=f(\alpha _{1},\alpha _{1},\alpha _{3},\ldots
,\alpha _{n})=f(0,\ldots ,0)$.
\hfill ~~~ \end{proof}

\begin{theorem}~\label{t3.1} Let
 $f\in P_k^n$, be a function which depends essentially on all of its $n$ variables
and $2\leq n\leq k$. Then  $f\in G_{n,k}^n$ if and
only if it can be represented as follows
\begin{equation}~\label{eq3.1}
f=[\bigoplus_{\beta_1\ldots\beta_n\notin Eq_k^n}
a_r.x_1^{\beta_1}\ldots x_n^{\beta_n}]\oplus a_0.[\bigoplus_{\alpha_1\ldots\alpha_n\in
Eq_k^n}x_1^{\alpha_1}\ldots x_n^{\alpha_n}],\end{equation} where
$r=\o{\beta_1\ldots\beta_n}$
 and at least two
among  the coefficients $$\{a_0\}\cup\{a_{r}\ |\
r=\o{\beta_1\ldots\beta_n},\  \& \ \beta_1\ldots\beta_n\notin
Eq_k^n\}$$
 are distinct.
 \end{theorem}
\begin{proof} $"\Ra"$
Let $f\in G_{n,k}^n$ be represented in its SC-form as follows
$$f=\bigoplus_{m=0}^{k^{n-1}}
a_m.x_1^{\beta_1}\ldots x_n^{\beta_n}\ \ \mbox{  where  }\ \ m=\o{\beta_1\ldots\beta_n}.$$

By
Lemma \ref{c33.1} we have $f(\alpha _{1},\ldots ,\alpha _{n})=f(0,\ldots ,0)=a_{0}$ for $%
\alpha _{1}\ldots \alpha _{n}\in Eq_{k}^{n}$. This shows that $f$ has to be
in the form (3.1). Moreover, at least two among the coefficients $%
\{a_{0}\}\cup \{a_{r}\mid r=\overline{\beta _{1}\ldots \beta _{n}}$ \& $%
\beta _{1}\ldots \beta _{n}\in Eq_{k}^{n}\}$ are distinct.

$"\La"$ Let  $f$     be represented in the form (\ref{eq3.1}). Then $f$ depends essentially on all of its variables since at least two among the coefficients
$\{a_0\}\cup\{a_{r}\ |\
r=\o{\beta_1\ldots\beta_n},\  \& \ \beta_1\ldots\beta_n\notin
Eq_k^n\}$,
 are distinct.
 We have to
prove that for all $i,j$ with  $1\leq j,i\leq n$ and $i\neq j$ the functions
$$f_{i\laa j}=a_0.\bigoplus_{\beta_i=\beta_j}
x_1^{\beta_1}\ldots x_j^{\beta_j}\ldots x_i^{\beta_i}\ldots x_n^{\beta_n}$$
do not depend on any of their variables.

 Without loss of generality let us prove this for  $j=1$ and $i=2$.
Then
$$f_{2\laa 1}= a_0.\big({x_1^0\oplus x_1^1\oplus\ldots\oplus x_1^{k-1}}\big)
\big(\bigoplus_{\beta_3\ldots\beta_n\in K^{n-2}}
x_3^{\beta_3}\ldots  x_n^{\beta_n})=a_0.1.1=a_0.$$
Hence $ess(f_{2\laa 1})=0$.
This completes the proof of the theorem.
 \hfill ~~~ \end{proof}

\begin{corollary}~\label{c3.2}
For each  $k,\ k\geq 3$ the functions
\begin{equation}~\label{eq3.3}p_k(x_1,\ldots,x_k)=
\bigoplus_{\alpha_1\ldots\alpha_k\notin Eq_k^k}
a_m.x_1^{\alpha_1}\ldots x_k^{\alpha_k},\end{equation} with
$a_m\in K$,
 have the essential arity gap
 $k$, excluding when all  $a_m$'s  are equal to $0$ and $p_k(x_i=x_j)=0$ for all $1\leq i,j\leq n$ with $i\neq j$.
\end{corollary}

\begin{theorem}~~\label{t3.2}
If $2\leq n\leq k$ then $$|G_{n,k}^n|=k^{\big({k\choose n}.n!+1\big)}-k.$$
\end{theorem}
\begin{proof}
The number of  coefficients $a_i$ in (\ref{eq3.1}) is equal to  ${k\choose n}.n!+1$ and they can be chosen
 in $k^{[{k\choose n}.n!+1]}$ ways.
There are $k$ "forbidden" cases, when
$a_0$ and $a_r$'s in  (\ref{eq3.1}) are the same.

 \hfill ~~~ \end{proof}

\begin{lemma}~\label{l3.3}
Let $f\in P_k^n$ be a $k$-valued function. If
$x_i\notin Ess(f_{u\laa v})$, with $1\leq i,u,v \leq n$, $u\neq v$ and $i\notin \{u,v\}$ then
$f_{u\laa v}=[f_{i\laa j}]_{u\laa v}$,  for all  $j$,
  $j\in \{1\ldots n\}$, $j\neq i$.
 \end{lemma}
 \begin{proof}\
Suppose with no  loss of generality  that $u=2$, $v=1$ and $i=3$.
Then
$f_{2\laa 1}=f(x_1,x_1,x_3,x_4,\ldots,x_n)$ and from $x_3\notin Ess(f_{2\laa 1})$ we have
$$f(x_1,x_1,\alpha,x_4,\ldots,x_n)=f(x_1,x_1,\beta,x_4,\ldots,x_n)$$ for all $\alpha,\beta\in K$.

Let $j\in\{1,\ldots,n\}$ and   $j\neq 3$. Then we have
  $$[f_{3\laa j}]_{2\laa 1}=f(x_1,x_1,x_j,x_4,\ldots,x_n)=f(x_1,x_1,x_3,x_4,\ldots,x_n)=f_{2\laa 1}.$$
  \hfill ~~~ \end{proof}

\begin{lemma}~\label{l3.4}
Let $f\in P_k^n$ be a $k$-valued function. If $x_v\notin Ess(f_{u\laa v})$
for some $ u,v \leq n$, then
$f_{u\laa v}=[f_{v\laa j}]_{u\laa j}=[f_{u\laa j}]_{v\laa j}$, for all  $j$,
  $j\in \{1\ldots n\}$, $j\neq u,v$.
 \end{lemma}
 \begin{proof}\
Without loss of generality assume that $u=2$ and  $v=1$.

Then
$f_{2\laa 1}=f(x_1,x_1,x_3,\ldots,x_n)$ and from $x_1\notin Ess(f_{2\laa 1})$ we obtain
$$f_{2\laa 1}=f(\alpha,\alpha,x_3,\ldots,x_n)=f(\beta,\beta,x_3,\ldots,x_n)$$ for
all $\alpha,\beta\in K$.

Let $j>2$ and  by symmetry, we may assume  $j=3$.
Then we obtain
  $$[f_{1\laa 3}]_{2\laa 3}=[f(x_3,x_2,x_3,x_4,\ldots,x_n)]_{2\laa 3}=f(x_3,x_3,x_3,x_4,\ldots,x_n)=$$
   $$=f(\alpha,\alpha,x_3,\ldots,x_n)=f_{2\laa 1}.$$
  \hfill ~~~ \end{proof}

 We are going to  describe the basic properties of the functions $f$ whose arity gap is non-trivial i.e.  $gap(f)=p$ with $2\leq p< n\leq k$, in the rest of the paper.

\begin{theorem}~\label{l3.5}
Let $2<p<n\leq k$. Then for each $f\in G_{p,k}^{n,}$ there is
a function $h\in P_{k}^{n}$ with \newline
(i) $ess(h)=n-p$;\newline
(ii) $f_{i\leftarrow j}=h$ for all $1\leq i,j\leq n$ with $i\neq j$ and $%
x_{i}\in Fic(h)$.\newline
Moreover, \ for all $1\leq u,v\leq n$ with $v\neq u$ and $x_{v}\in
Fic(f_{u\leftarrow v})$ holds $f_{i\leftarrow j}=f_{u\leftarrow v}$ for all $%
1\leq i,j\leq n$ with $i\neq j$ and $x_{i}\in Fic(f_{u\leftarrow v})$ as well as $ess(f_{u\leftarrow v})=n-p$.
\end{theorem}

\begin{proof}
Let $f\in G_{p,k}^{n}$ and $1\leq i,j\leq n$ with $i\neq j$ and $%
ess(f_{i\leftarrow j})=n-p$. Let us set $h=f_{i\laa j}$.

First, we shall prove that there are $1\leq r,s\leq n$ with $r\neq s$, $ess(f_{r\leftarrow s})=n-p$ and $x_{s}\notin
 Ess(f_{r\rightarrow s})$.

 If $x_{j}\notin Ess(f_{i\leftarrow j})$  we are done in this part of the proof.

Furthermore, let us assume that $x_{j}\in Ess(f_{i\leftarrow j})$.
Since $n>p>2$,  i.e. $n>3$ there are $1\leq r,s\leq n$ with $r\neq s$ and $x_{r},x_{s}\in
Fic(h)\setminus \{x_{i}\}$. We shall prove that $ess(f_{r\leftarrow s})=n-p$ and $x_{s}\notin
 Ess(f_{r\rightarrow s})$.

 By Lemma \ref{l3.3}, we have $$f_{i\leftarrow
j}=[f_{r\leftarrow s}]_{i\leftarrow j}.$$ This gives $n-p=ess(f_{i\leftarrow
j})\leq ess(f_{r\leftarrow s})\leq n-p$ since $gap(f)=p$, i.e. $%
ess(f_{r\leftarrow s})=n-p$. Further, let $x_{m}\in Ess(h)\setminus \{x_{j}\}
$. Assume that $x_{m}\notin Ess(f_{r\rightarrow s})$. Then
 $$f_{r\leftarrow
s}=[f_{r\leftarrow s}]_{m\leftarrow i}\quad\mbox{ and thus }\quad [f_{r\leftarrow
s}]_{i\leftarrow j}=[[f_{r\leftarrow s}]_{m\leftarrow i}]_{i\leftarrow j}
.$$
Moreover, $h=h_{r\leftarrow s}$ since $x_{r}\notin Ess(h)$. Because of $%
r\neq i$, we have
$$h_{r\leftarrow s}=[f_{r\leftarrow s}]_{i\leftarrow j}\quad\mbox{ and thus }\quad h=[[f_{r\leftarrow s}]_{m\leftarrow i}]_{i\leftarrow j}.$$
Since $x_{m}\neq x_{j}$ we have $x_{m}\notin Ess([[f_{r\leftarrow
s}]_{m\leftarrow i}]_{i\leftarrow j})$, i.e. $x_{m}\notin Ess(h)$, which is a
contradiction. This shows that
$$Ess(h)\setminus \{x_{j}\}\subseteq
Ess(f_{r\leftarrow s})\quad\mbox{ and thus }\quad Fic(f_{r\leftarrow s})\subseteq
\{x_{j}\}\cup Fic(h).$$

Now let us  prove that $x_{s}\notin Ess(f_{r\leftarrow s})$.
 For suppose this were not true i.e.  $x_{s}\in Ess(f_{r\leftarrow s})$.
Because of $n-p=ess(f_{r\leftarrow s})=ess(f_{i\leftarrow j})$, then $x_{s}\notin
Fic(f_{r\leftarrow s})$ implies $Fic(f_{r\leftarrow s})=(\{x_{j}\}\cup
Fic(h))\setminus \{x_{s}\}$. This provides $x_{i},x_{j}\in
Fic(f_{r\leftarrow s})$. Then we have $f_{r\leftarrow s}=[f_{r\leftarrow
s}]_{i\leftarrow j}$. Moreover, $h=h_{r\leftarrow s}$. Since $r\neq i,$ we
have $[f_{r\leftarrow s}]_{i\leftarrow j}=h_{r\leftarrow s}$. Altogether,
this provides $f_{r\leftarrow s}=h$. This shows that $Fic(h)=Fic(f_{r%
\leftarrow s})$, and in particular, $x_{j}\in Fic(h)$ which is a contradiction. Hence $x_{s}\notin Ess(f_{r\leftarrow s})$.

Second, we shall prove that $h=f_{u\leftarrow v}$ for all $%
1\leq u,v\leq n$ with $u\neq v$ and $x_{u}\in Fic(h)$.

Let $1\leq r,s\leq n$ with $r\neq s$ and $%
x_{r}\in Fic(h)$. Assume that $r\neq i$. By the same arguments as in the
previous, we can show that $Ess(h)\subseteq Ess(f_{r\leftarrow s})$. Because
of $gap(f)=p$, this implies $Ess(h)=Ess(f_{r\leftarrow s})$. Let us assume  that $%
s\neq i$. Since $x_{r}\notin Ess(h)$ and $x_{i}\notin
Ess(h)=Ess(f_{r\leftarrow s})$, we have
 $$h=h_{r\leftarrow s}\quad\mbox{ and }\quad
f_{r\leftarrow s}=[f_{r\leftarrow s}]_{i\leftarrow l}$$
 for $l\in \{j,s\}$.
Since $s\neq i$ and $r\neq i$, it is easy to see that $h_{r\leftarrow
s}=[f_{r\leftarrow s}]_{i\leftarrow j}$ when $r\neq j$ and $h_{r\leftarrow
s}=[f_{r\leftarrow s}]_{i\leftarrow s}$ when $r=j$. Altogether, this gives $%
f_{r\leftarrow s}=h$.

Next, let us assume  that $s=i$. Since $x_{j}\notin
Ess(h)=Ess(f_{r\leftarrow s})$, we have
$f_{r\leftarrow s}=[f_{r\leftarrow
s}]_{j\leftarrow s}$. Moreover, $x_{j},x_{r}\notin Ess(h)$ implies $%
h=[h_{r\leftarrow s}]_{j\leftarrow s}$. Since $s=i$, we have
$$
[f_{r\leftarrow s}]_{j\leftarrow s}=[h_{r\leftarrow s}]_{j\leftarrow s}\quad\mbox{ and
thus }\quad f_{r\leftarrow s}=h.$$

 Finally, assume  that $r=i$ and $s\neq i$. Because of $x_{j}\notin
Ess(h)$, we have in particular $h=f_{j\leftarrow i}$ where $x_{i}\notin
Ess(h)=Ess(f_{j\leftarrow i})$ (as we have shown in the previous). We might
choose $\widetilde{h}:=f_{j\leftarrow i}$ (instead of $h:=f_{i\leftarrow j}$%
) and obtain $f_{j\leftarrow i}=f_{i\leftarrow s}$ by the previous
considerations. Altogether, we have $h=f_{i\leftarrow s}$.

  \hfill ~~~ \end{proof}

\begin{lemma}~~\label{t3.5}
Let $f\in G_{p,k}^n$ .  Then the following conditions hold:

$(i)$ If $2<p<n$ then there exist $u,v\in \{1,\ldots,n\}$
such that $f_{u\laa v}$ depends
essentially on $n-p$ variables and $x_v\in Ess(f_{u\laa v})$;

$(ii)$ If $2<p\leq n$ then there exist $u,v\in \{1,\ldots,n\}$ such that $f_{u\laa v}$ depends
 essentially on $n-p$ variables and $x_v\notin Ess(f_{u\laa v})$.
\end{lemma}
\begin{proof}
By Theorem \ref{l3.5}, there are $u,v\in \{1,\ldots ,n\}$ such that $%
ess(f_{u\leftarrow v})=n-p$ and $f_{u\leftarrow v}=f_{i\leftarrow j}$ for $%
1\leq i,j\leq n$ with $i\neq j$ and $x_{i}\in Fic(f_{u\leftarrow v})$.

(i) Since $p<n$, there is an $l\in \{1,\ldots ,n\}$ with $x_{l}\in
Ess(f_{u\leftarrow v})$. Then $f_{u\leftarrow v}=f_{u\leftarrow l}$ since $%
x_{u}\in Fic(f_{u\leftarrow v})$. This shows $x_{l}\in Ess(f_{u\leftarrow l})
$ where $ess(f_{u\leftarrow l})=n-p$.

(ii) It was already proved in Theorem \ref{l3.5}.
 \hfill ~~~ \end{proof}

\begin{theorem}~\label{t3.6}
Let $f$ be a $k$-valued function
which depends essentially on the all of its  $n$ variables and
$2<p< n\leq k$. Then $f\in G_{p,k}^n$ if and only if  there exist $n-p$ variables $y_{l_1},\ldots,y_{l_{n-p}}\in X_n$ such that
\begin{equation}\label{eq3.6}
f=h \oplus g,
\end{equation}
where $ Ess(h)=\{y_{l_1},\ldots,y_{l_{n-p}}\}$
and $g\in G_{n,k}^n$. 
Moreover,  $g_{i\leftarrow j}=0$ for all $%
1\leq i,j\leq n$ with $i\neq j$.

\end{theorem}
\begin{proof}
$"\La"$ Let $f$ be represented in the form  (\ref{eq3.6}),
where $h$ depends essentially
on all of its $n-p$ variables and $g\in G_{n,k}^n$. With no loss of generality we might assume that $h$ is an $(n-p)$-ary $k$-valued function.

Again, without loss of generality, let us assume that $y_{l_1},\ldots,y_{l_{n-p}}=x_{1},\ldots,x_{{n-p}}$.

Since  $2<p< n\leq k$  there is at least one variable $x_j\in X_n$
on which $h$
does not depend essentially i.e. $n-p<j\leq n$.
Then from Lemma \ref{l2.3} we have
$Ess(h_{j\laa i})=Ess(h)$ for all $i,$ $1\leq i\leq n-p$. Since $g\in G_{n,k}^n$ it follows that
 $ess(g_{u\laa v})=0$
 for all $u$ and $v$ with  $1\leq u,v\leq n$ and $u\neq v$.
Hence  $ess(f_{u\laa v})=ess(h_{u\laa v})$ for all $u$, $v$  with  $1\leq u,v\leq n$ and $u\neq v$.
On the other hand $ess(h_{u\laa v})=n-p$ when $u>n-p$.
  Consequently, $$\max_{t\in Min(f)}ess(t)=n-p\quad\mbox{and\ hence}\quad gap(f)=p.$$

 ${"\Ra"}$
From Lemma \ref{t3.5} $(ii)$ it follows that there are  $u,v\in \{1,\ldots,n\}$ such that $f_{u\laa v}$ depends
 essentially on $n-p$ variables and $x_v\notin Ess(f_{u\laa v})$. With no loss of generality let us assume that $(v,u)=(n-1,n)$ and let  $h:=f_{n\laa n-1}$, where  $Ess(h)=\{x_1,\ldots,x_{n-p}\}$.

 Let $g\in P_k^n$ be the function defined by
\begin{equation}\label{eq3.5}
g:=f\ominus h
\end{equation}
i.e. $g$ is the unique function in $P_k^n$ such that
$g\oplus h=f$.

We have to prove that $g\in G_{n,k}^n$ i.e. $Ess(g)=X_n$ and  $g_{i\laa j} =0$ for all $i,j$ with
$1\leq j,i\leq n$ and $i\neq j$.

First, $\{x_{n-p+1},\ldots,x_n\}\subseteq Ess(g)$ because $Ess(f)=X_n$ and  $\{x_{n-p+1},\ldots,x_n\}\cap Ess(h)=\emptyset$.

Second, to prove that $\{x_1,\ldots,x_{n-p}\}\subseteq Ess(g)$ we shall suppose that this is not the case and without loss of generality assume that $x_1\notin Ess(g)$. Then from Theorem \ref{l3.5} we obtain
$$f_{n\laa 1}=h(x_1,x_2,\ldots,x_{n-p})\oplus g_{n\laa 1}=h(x_1,x_2,\ldots,x_{n-p}).$$
Since $K$ is an additive group it follows that $g_{n\laa 1}=0$. Consequently $g_{1\laa n}=0$ and from Lemma \ref{l2.3} we obtain $g=g_{1\laa n}=0$ which is a contradiction. Hence  $\{x_1,\ldots,x_{n-p}\}\subseteq Ess(g)$ and $Ess(g)=X_n$.

To prove that  $ess(g_{i\laa j}) =0$ for all $i,j$ with
$1\leq i,j\leq n$ and $i\neq j$, we shall consider several cases.

{\bf Case 1.} Let $n-p<j,i<n$ and $i\neq j$. According to (\ref{eq3.5}) we have
$g_{n\laa n-1}=f_{n\laa n-1}\ominus h=0$.
From Theorem \ref{l3.5}, it follows that $f_{i\laa j}=f_{n\laa n-1}$ and from Lemma \ref{l2.3} we have  $h_{i\laa j}=h$ which implies  $g_{i\laa j}=0$ i.e. $ess(g_{i\laa j})=0$ for all
$i,j$ with $n-p<j,i<n$ and $i\neq j$.

 {\bf Case 2.}  Let $j\leq n-p<i$. We may assume that $j=1$
 and $i=n$.
 Since $x_{n}\notin
Ess(h)$, we have $h=f_{n\leftarrow n-1}=f_{n\leftarrow 1}$ by Theorem \ref{l3.5}.
Moreover, from Lemma \ref{l2.3} we have $h=h_{n\leftarrow 1}$.  Thus
$g_{n\leftarrow 1}=f_{n\leftarrow
1}\ominus h_{n\leftarrow 1}=h\ominus h=0.$
 On the other hand
we have
$$h_{1\leftarrow
n}=[f_{n\leftarrow n-1}]_{1\leftarrow
n}=[f_{n\leftarrow 1}]_{1\leftarrow n}=f_{1\leftarrow n},$$
 i.e. $h_{1\leftarrow
n}=f_{1\leftarrow n}$ and thus $g_{1\leftarrow n}=f_{1\leftarrow n}\ominus h_{1\leftarrow
n}=0$. Hence $g_{1\leftarrow n}=g_{n\leftarrow 1}=0$.

{\bf Case 3.} Let $i,j\leq n-p$ and $i\neq j$. By symmetry we may assume that $j=1$ and $i=2$.
We have to prove that
$g_{2\laa 1}=0$.

 First let us assume that  there is some $u>n-p$ such
that $x_u\notin Ess(f_{2\laa 1})$.  According to Case 2 we have $$g_{u\leftarrow 1}=g_{u\leftarrow 2}=g_{1\leftarrow u}=g_{2\leftarrow u}=0.$$
From Lemma \ref{l3.4} we obtain $g_{2\leftarrow u}=[g_{2\leftarrow 1}]_{u\leftarrow 1}$ and  Lemma  \ref{l2.3} implies
$g_{2\leftarrow 1}=[g_{2\leftarrow 1}]_{u\leftarrow 1}$. Hence $g_{2\leftarrow u}=g_{2\leftarrow 1}=0$.

Second, assume that $\{x_{n-p+1},\ldots,x_n\}\subseteq Ess(f_{2\laa 1})$.
Since  $p\geq 3$ then $f\in G_{p,k}^n$ implies $ess(f_{2\laa 1})<n-2$.

Hence in the whole Case 3 the function   $f_{2\laa 1}$ can
   essentially depend on at most
$n-3$ variables. Then  there is a variable $x_v$, $3\leq v\leq n-p$,
 which is not essential in $f_{2\laa 1}$. Without loss of generality
  let us assume that
$v=3$. 
From Theorem \ref{l3.5} we have $f_{n-1\leftarrow 3}=h$ and hence 
$f_{3\leftarrow n-1}=h_{3\leftarrow n-1}$. This implies
 $ 
[f_{3\leftarrow n-1}]_{2\leftarrow 1}=[h_{3\leftarrow n-1}]_{2\leftarrow 1}$. 
Hence  $[f_{3\leftarrow n-1}]_{2\leftarrow 1}=[f_{2\leftarrow
1}]_{3\leftarrow n-1}=f_{2\leftarrow 1}$ and $[h_{3\leftarrow
n-1}]_{2\leftarrow 1}=[[f_{2\leftarrow 1}]_{3\leftarrow n-1}]_{n\leftarrow
n-1}$. From $f_{2\leftarrow 1}=[f_{2\leftarrow 1}]_{3\leftarrow n-1}$ it
follows $$[h_{3\leftarrow n-1}]_{2\leftarrow 1}=[f_{2\leftarrow
1}]_{n\leftarrow n-1}=[f_{n\leftarrow n-1}]_{2\leftarrow 1}=h_{2\leftarrow
1}.$$  Altogether, we have $f_{2\leftarrow 1}=h_{2\leftarrow 1}$, i.e. $%
g_{2\leftarrow 1}=0$ and $ess(g_{2\laa 1})=0$.

 Consequently, $g\in G_{n,k}^n$ and $g_{i\leftarrow j}=0$ for all 
$1\leq i,j\leq n$ with $i\neq j$.

 \hfill ~~~ \end{proof}

\begin{corollary}~~\label{r3.1} Let $f\in G_{p,k}^n$. Then there is a partition of the set $Ess(f)=\{x_1,\ldots,x_n\}$ into the sets
 $V:=Ess(h)$ and $W:=Ess(f)\setminus V$, where $h$ is the function defined in the proof of Theorem \ref{t3.6}, such that

$$(x_i,x_j)\in  W^2\Rightarrow   \big(ess(f_{i\laa j})=n-p\
\& \  x_j\notin Ess(f_{i\laa j})\big)$$ and
$$(x_i, x_j)\in W\times V\Rightarrow \big(ess(f_{i\laa j})=n-p\
\&\  x_j\in Ess(f_{i\laa j})\big).$$

\end{corollary}

\begin{theorem}~\label{t3.7}
If $2< p< n\leq k$,  then
$$ |G_{p,k}^n|= \big(k^{{k\choose n}.n!}-1\big).\sum_{j=p}^n(-1)^{j-p}{j\choose p}{n\choose j}.k^{k^{n-j}} .$$
\end{theorem}
\begin{proof}
Let  $f=h\oplus g\in G_{p,k}^n$ and $g_{i\leftarrow j}=0$ for all 
$1\leq i,j\leq n$ with $i\neq j$. Let $a\in K$ be a non-zero natural number from $K$ i.e. $0<a\leq k-1$. Then clearly $t=g\oplus a\in G_{n,k}^n$ and $t_{i\leftarrow j}=a$ for all 
$1\leq i,j\leq n$ with $i\neq j$. According to Theorem \ref{t3.2} the number of functions $g\in G_{n,k}^n$ with  $g_{i\leftarrow j}=0$ for all 
$1\leq i,j\leq n$ when $i\neq j$ is equal to  $ |G_{n,k}^n|/k= \big(k^{{k\choose n}.n!}-1\big)$.

It is well known that the number of all functions in $P_k^n$ which depend essentially on exactly $n-p$ variables is equal to $|P_{n-p,k}^n|=\sum_{j=p}^n(-1)^{j-p}{j\choose p}{n\choose j}.k^{k^{n-j}} $.

Consequently,
$$ |G_{p,k}^n|:=(|G_{n,k}^n|/k).(|P_{n-p,k}^n|) = \big(k^{{k\choose n}.n!}-1\big).\sum_{j=p}^n(-1)^{j-p}{j\choose p}{n\choose j}.k^{k^{n-j}} .$$

 \hfill ~~~ \end{proof}

\section{The class $G_{2,k}^n$ with $4\leq n\leq k$}\label{sec6}
~~\\

There are two subclasses of the 
 class $G_{2,k}^n$ with $4\leq n\leq k$. First one, consists of functions satisfying conditions similar  as the conditions in Theorem \ref{l3.5}, Lemma \ref{t3.5} and Theorem \ref{t3.6}  for $p=2$. The second subclass consists of functions whose behavior is similar to the functions from $G_{2,k}^n$ with $n>k$ (Theorem 2.1 in \cite{ros}).

 \begin{lemma}~\label{th31.9}
Let $f$ be a $k$-valued  function which depends essentially on all of its $n$, $n>3$ variables and $gap(f)=2$. Then there exist   two distinct essential variables $x_u,x_v$ such that  $ess(f_{u\laa v})=n-2$, and $x_v\notin Ess(f_{u\laa v})$.
Moreover, $ess(f_{u\laa m})=ess(f_{v\laa m})=n-2$ for all $m$, $1\leq m\leq n$ with $m\notin \{u,v\}$.
\end{lemma}
\begin{proof}
Since $gap(f)=2$, there are $1\leq i,j\leq n$ with $i\neq
j$ and $ess(f_{i\leftarrow j})=n-2$.

If $x_{j}\notin Ess(f_{i\leftarrow j})$ we are done.

Let us assume that $x_{j}\in Ess(f_{i\leftarrow j})$.
 Since $gap(f)=2$, there is
a $w\in \{1,\ldots ,n\}\setminus \{i,j\}$ such that $x_{w}\notin
Ess(f_{i\leftarrow j})$. From Lemma \ref{l3.3}, we obtain
 $$[f_{w\leftarrow
j}]_{i\leftarrow j}=f_{i\leftarrow j}\quad\mbox{ and }\quad n-2=ess(f_{i\leftarrow j})\leq
ess(f_{w\leftarrow j}).$$
 Hence $ ess(f_{i\leftarrow j})=ess(f_{w\leftarrow j})=n-2$
 since $gap(f)=2$.

We shall prove that   $x_{j}\notin Ess(f_{w\leftarrow j})$ or  $x_{i}\notin Ess(f_{w\leftarrow i})$ which will complete the proof.

For suppose this were not true.
 Then $x_{j}\in Ess(f_{w\leftarrow j})$ and $x_{i}\in Ess(f_{w\leftarrow i})$. Assume that $%
x_{i}\in Ess(f_{w\leftarrow j})$. Then  there is a $r\in \{1,\ldots
,n\}\setminus \{i,j,w\}$ such that $x_{r}\notin Ess(f_{w\leftarrow j})$.
Then $x_{r}\notin Ess([f_{w\leftarrow j}]_{i\leftarrow j})$, i.e. $%
x_{r}\notin Ess(f_{i\leftarrow j})$, which is a contradiction. Hence $x_{i}\notin
Ess(f_{w\leftarrow j})$. By similar arguments, we obtain that $x_{j}\notin
Ess(f_{w\leftarrow i})$.

Thus we have
$Ess(f_{i\laa j})=X_n\setminus\{x_i,x_w\},$ $Ess(f_{w\laa i})=X_n\setminus\{x_j,x_w\}$  and $ Ess(f_{w\laa j})=X_n\setminus\{x_i,x_w\}.$

Since $n>3$ it follows that there is at least one essential variable $x_s$ in $f$ with $s\in \{1,\ldots
,n\}\setminus \{i,j,w\}$.

Now, the equation $Ess(f_{w\laa s})=X_n\setminus\{x_w\}$ is implicit in \cite{ros} (Lemma 1.1 (5)). This equation contradicts $gap(f)=2$.

Let $m\notin \{u,v\}$ be a natural number with $1\leq m\leq n$. Lemma \ref{l3.3} implies 
$f_{u\leftarrow
v}=[f_{u\leftarrow m}]_{v\leftarrow m}$
for all  $1\leq m\leq n$ with $m\notin \{u,v\}$. Hence $n-2= ess(f_{u\leftarrow v})=ess([f_{u\leftarrow m}]_{v\leftarrow m})\leq ess(f_{u\leftarrow m})$. Now, $gap(f)=2$ shows that $ess(f_{u\leftarrow m})=n-2$ and by symmetry we obtain
$ess(f_{v\leftarrow m})=n-2$ 
\hfill ~~~\end{proof}

Let us denote by $G_{p,k}^{n,+}$ the set of all functions $f\in G_{p,k}^{n}$ for which there exist $i$ and $j$ with $1\leq i,j\leq n$ such that $i\neq j$, $x_j\in Ess(f_{i\laa j})$ and $ess(f_{i\leftarrow j})=n-p$.

 $G_{p,k}^{n,-}$ denotes  the set of all functions $f\in G_{p,k}^{n}$ for which  $x_v\notin Ess(f_{u\laa v})$  for all $1\leq u,v\leq n$ with $u\neq v$.
 
 \begin{proposition}\label{pr5.1}
 If $3<n\leq k$ then $G_{2,k}^{n}=G_{2,k}^{n,+}\cup  G_{2,k}^{n,-}$. 
\end{proposition}
\begin{proof}
Clearly, $G_{2,k}^{n,+}\cup  G_{2,k}^{n,-}\subseteq G_{2,k}^{n}$. Let $f\in G_{2,k}^{n}$. Then    
 Lemma \ref{th31.9} implies  that there exist two distinct essential variables $x_u,x_v$ such that  $ess(f_{u\laa v})=n-2$, and $x_v\notin Ess(f_{u\laa v})$.
 
  If $x_j\notin Ess(f_{i\laa j})$ for all $1\leq i,j\leq n$ with $i\neq j$ then $f\in G_{2,k}^{n,-}$.
  
  Next, assume that  there are $1\leq i,j\leq n$ with  $x_j\in Ess(f_{i\laa j})$. We have to prove that   
  $f\in G_{2,k}^{n,+}$ i.e. there exist  $r,s$ with $1\leq r,s\leq n$,  $x_r\in Ess(f_{s\laa r})$ and $ess(f_{s\laa r})=n-2$.
  
  If $\{i,j\}\cap\{u,v\}\neq\emptyset$ we are done because of Lemma \ref{th31.9}.
  
  Let  $\{i,j\}\cap\{u,v\}=\emptyset$. From Lemma \ref{th31.9} we have  $ess(f_{u\laa i})=ess(f_{u\laa j})=ess(f_{v\laa i})=ess(f_{v\laa j})=n-2$. If $x_j\in Ess(f_{u\laa j})$ we are done as above. If $x_j\notin Ess(f_{i\laa j})$ then we have $x_i\in Ess(f_{i\laa j})$ and $ess(f_{u\laa i})=n-2$.
 \end{proof}
\begin{corollary}
$$G_{p,k}^{n}=\left\{\begin{array}{llll}
G_{p,k}^{n,+} &if  &2<p<n\leq k  &  \\ 
 &  &  &  \\ 
 G_{p,k}^{n,-}& if & (2\leq n\leq k\  \&\  p=n) & or\ (n>k )\\ 
 &  &  &  \\ 
G_{p,k}^{n,+}\cup  G_{p,k}^{n,-} & if  & 3<n\leq k\ \&\ p=2. & 
\end{array} \right.
$$
\end{corollary}
\begin{proof}
This representation of the set $G_{p,k}^{n}$ follows by 
 Theorem \ref{t3.6}, Theorem \ref{t3.1}, Proposition  \ref{pr5.1}, Theorem 3.1 in \cite{s2} and Theorem 2.1 in \cite{ros}.  
\end{proof}~~~
\\
\\

\subsection{The subclass $G_{2,k}^{n,+}$}
~~~\\
\\
The first representation is related to Theorem 16 in \cite{mig1}.

\begin{theorem}~\label{th31.7}
Let $4\leq n\leq k$ and $f\in P_{k}^{n}$. Then the
following statements are equivalent:

(i) $f\in G_{2,k}^{n,+}$;

(ii) There is a function  $h\in P_{k}^{n}$ with $ess(h)=n-2$ and $%
f_{r\leftarrow s}=h$ for all $1\leq r,s\leq n$ with $r\neq s$ and $x_{r}\in
Fic(h)$.
\end{theorem}
\begin{proof}
$(ii)\Rightarrow (i)$ is clear.

$(i)\Rightarrow (ii)$. Let $i$ and $j$ be two distinct natural numbers fro which $1\leq i,j\leq n$, $x_j\in Ess(f_{i\laa j})$ and $ess(f_{i\laa j})=n-2$.  Following the proof of
 Lemma \ref{th31.9} we might conclude that  there exists an essential variable  $x_w$  in $f$ such that   $x_w\notin Ess(f_{i\laa j})$,  $ess(f_{w\laa i})=ess(f_{w\laa j})=n-2$, and $x_i\notin Ess(f_{w\laa i})$ or $x_j\notin Ess(f_{w\laa j})$. Without loss of generality let us assume that  $x_i\notin Ess(f_{w\laa i})$ i.e. $Ess(f_{w\laa  i})=X_n\setminus\{x_w,x_i\}$.

From Lemma \ref{th31.9} we have
 $ess(f_{w\leftarrow r})=ess(f_{i\leftarrow r})=n-2$ for all $r$, $1\leq r\leq n$ with $r\notin \{w,i\}$.

First, we shall show that $X_n\setminus\{x_w,x_i,x_r\}\subset Ess(f_{w\laa r})$ and $X_n\setminus\{x_w,x_i,x_r\}\subset Ess(f_{i\laa r})$ for all $r$, $1\leq r\leq n$ with $r\notin \{w,i\}$. Note that $X_n\setminus\{x_w,x_i,x_r\}\neq\emptyset$ because $n>3$. Since $x_i\notin Ess(f_{w\laa i})$ and from Lemma  \ref{l3.4} it follows
$$f_{w\laa i}=[f_{w\laa r}]_{i\laa r}=[f_{w\laa i}]_{i\laa r}.$$
Let $x_s$ be an essential variable in $f$ with $s\notin\{w,i,r\}$.
 Suppose that $x_s\notin  Ess(f_{w\laa r})$.
Hence  $x_s\notin  Ess([f_{w\laa i}]_{i\laa r})$ i.e. $x_s\notin  Ess(f_{w\laa i})$, which contradicts  $Ess(f_{w\laa  i})=X_n\setminus\{x_w,x_i\}$.

Second, we shall prove that $x_r\in  Ess(f_{w\laa r})$ and $x_r\in  Ess(f_{i\laa r})$ for all $r$, $1\leq r\leq n$ with $r\notin \{w,i\}$.
Since $x_{j}\in Ess(f_{i\leftarrow j})$ it follows that $j\notin \{w,i\}$.

Consider the case $r=j$. We have known that $x_{j}\in Ess(f_{i\leftarrow j})$ and from  $x_w\notin Ess(f_{i\laa j})$ we obtain
$$f_{i\laa j}=[f_{i\laa j}]_{w\laa j}=[f_{i\laa w}]_{w\laa j}=f_{i\laa w}.$$
Hence  $x_{j}\in Ess(f_{w\leftarrow j})$.

Assume that $r\notin \{w,i,j\}$. Suppose that $x_r\notin Ess(f_{i\laa r})$  and from Lemma \ref{l3.4} we have $f_{i\laa r}=[f_{i\laa j}]_{r\laa j}$. Since $x_w\notin Ess(f_{i\laa j})$ we have $x_w\notin Ess([f_{i\laa j}]_{r\laa j})$ i.e. $x_w\notin Ess(f_{i\laa r})$. Then $ess(f_{i\laa r})=n-3$ because $x_r,x_i,x_w\notin Ess(f_{i\laa r})$ which is a contradiction. Hence $x_r\in Ess(f_{i\laa r})$. In a similar way it might be shown that $x_r\in Ess(f_{w\laa r})$.

 Hence  $Ess(f_{i\laa r})=Ess(f_{w\laa r})= Ess(f_{w\laa i})$ for all $r$, $1\leq r\leq n$ with $r\notin \{w,i\}$.

 Finally, let us set $h:=f_{w\laa i}$. Clearly  $Ess(h)=X_n\setminus\{x_w,x_i\}$ and $ess(h)=n-2$. Let $r$ and $s$ be two natural numbers such that $1\leq r,s\leq n$, $r\neq s$ and $r\in\{w,i\}$. With no loss of generality let us assume that $r=w$. Since   $x_i\notin Ess(f_{r\laa i})$ then  Lemma \ref{l3.4} implies
 $$h=f_{r\laa i}=[f_{r\laa s}]_{i\laa s}=[f_{i\laa s}]_{r\laa s}=f_{r\laa s},$$
 because $x_i\notin Ess(f_{r\laa s})$.

 \hfill ~~~\end{proof}

 \begin{corollary}~\label{c3.8}
If  $f\in G_{2,k}^{n,+}$, $n>3$ then there exist $x_u,x_v\in Ess(f)$ such that
$f_{r\laa s}=f_{u\laa v}=h$ for all  $r\in \{u,v\}$ and  $s\in \{1,\ldots,n\}$, $s\neq r$, as well as $ess(f_{u\leftarrow v})=n-2$.
 \end{corollary}

 \begin{theorem}~\label{th31.8}
Let $f$ be a $k$-valued function
which depends essentially on the all of its  $n$ variables , $n>3$. Then the following sentences are equivalent:

$(i)$  $f\in G_{2,k}^{n,+}$;

$(ii)$ There exist $n-2$ variables $y_{l_1},\ldots,y_{l_{n-2}}\in X_n$ such that
$$
f=h \oplus g,
$$
where $Ess(h)=\{y_{l_1},\ldots,y_{l_{n-2}}\}$ 
and $g\in G_{n,k}^n$. Moreover   $g_{i\laa j}=0$ for all $1\leq i,j\leq n$ with $i\neq j$.
\end{theorem}
\begin{proof}
We  might prove the theorem in a similar way as  Theorem \ref{t3.6} by using Theorem \ref{th31.7} instead of Theorem \ref{l3.5}.
 \hfill ~~~\end{proof}

\begin{proposition}~\label{p5.3}
 $|G_{2,k}^{n,+}|= \big(k^{{k\choose n}.n!}-1\big).\sum_{j=2}^n(-1)^{j }{j\choose 2}{n\choose j}.k^{k^{n-j}} .$
 \end{proposition}

The proof can be done in a similar way as the proof of Theorem \ref{t3.7}.
~~\\ 
\subsection{The subclass $G_{2,k}^{n,-}$}
~~\\

Thus Lemma \ref{th31.9} implies  $G_{2,k}^{n,+}\cup G_{2,k}^{n,-}=G_{2,k}^{n}$.
We are going to describe the class $ G_{2,k}^{n,-}$ when $3<n\leq k$.

The next theorem is proved for $n>k$ by R. Willard in \cite{ros}.

 \begin{theorem}\label{t5.5}
Let $f$ be a $k$-valued  function which depends essentially on all of its $n$, $n>3$ variables.  If $f\in G_{2,k}^{n,-}$ then each identification minor of $f$ is a  symmetric function with respect to its essential variables.
 \end{theorem}
 \begin{proof}
Since $G_{2,k}^{n,-}\subset G_{2,k}^{n}$ then   Lemma \ref{th31.9} implies that  there  exist   two distinct essential variables $x_u,x_v$ such that  $ess(f_{u\laa v})=n-2$, and $x_v\notin Ess(f_{u\laa v})$. With no loss of generality let us assume that $(v,u)=(1,2)$ and   it is enough to prove that  $f_{2\laa 1}$ is a symmetric function with respect to the variables from the set $Ess(f_{2\laa 1})$. Since $Ess(f_{2\laa 1})=\{x_3,\ldots,x_n\}$ there is an $n-2$-ary function $h:K^{n-2}\to K$ such that  $f_{2\laa 1}=h(x_3,\ldots,x_n)$. By symmetry,  we have to prove that $$h(x_3,x_4,x_5,\ldots,x_n)=h(x_4,x_3,x_5,\ldots,x_n).$$

  In fact, we obtain

  \begin{tabular}{l}
  $h(x_3,x_4,x_5\ldots,x_n)$=\\
=$f(x_1,x_1,x_3,x_4,x_5,\ldots,x_n)$\\
=$f(x_3,x_3,x_3,x_4,x_5,\ldots,x_n)$\quad
(  since $x_1\notin Ess(f_{2\laa 1}$)\\
=$f(x_4,x_3,x_4,x_4,x_5,\ldots,x_n)$\quad
(  since $x_3\notin Ess(f_{1\laa 3}$)\\
=$f(x_4,x_3,x_3,x_3,x_5,\ldots,x_n)$\quad
(  since $x_4\notin Ess(f_{3\laa 4}$)\\
=$f(x_4,x_4,x_4,x_3,x_5,\ldots,x_n)$\quad
(  since $x_3\notin Ess(f_{2\laa 3}$)\\
=$f(x_1,x_1,x_4,x_3,x_5,\ldots,x_n)$\quad
(  since $x_1\notin Ess(f_{2\laa 1}$)\\
=$h(x_4,x_3,x_5,\ldots,x_n)$.
  \end{tabular}

  \hfill ~~~\end{proof}
We can now give a representation of the functions in $G_{2,k}^{n,-}$ $($see also Theorem 16 in \cite{mig1}$)$
 \begin{theorem}~\label{th5.6}
Let $f$ be an $n$-ary  $k$-valued function with $3<n\leq k$.
  Then the following sentences are equivalent:

$(i)$  $f\in G_{2,k}^{n,-}$;

$(ii)$ $f=t\oplus g$ where $g\in G_{n,k}^n$  and $t$ is an $n$-ary totally symmetric function with  $Ess(t_{i\laa j})=X_n\setminus\{x_i,x_j\}$ for all $i,j\in \{1,\ldots,n\}$, $i\neq j$. Moreover   $g_{i\laa j}=0$ for all $1\leq i,j\leq n$ with $i\neq j$. 
\end{theorem}
\begin{proof}
"$\La$" Clearly, $ess(t_{i\laa j})=n-2$ and $ess(g_{i\laa j})=0$  for all $i,j\in \{1,\ldots,n\}$, $i\neq j$. Hence $ess(f_{i\laa j})=n-2$ and $gap(f)=2$. We have to prove that  $Ess(f)=X_n$. By symmetry it is enough to show that $x_1\in Ess(t\oplus g)$.
Since $n>3$ we have
$$f(x_1,x_2,x_2,x_4,\ldots,x_n)=f_{3\laa 2}=t_{3\laa 2}\oplus g_{3\laa 2}$$
and $$Ess(f_{3\laa 2})=Ess(t_{3\laa 2}\oplus g_{3\laa 2})=\{x_1,x_4,\ldots,x_n\}=Ess(t_{3\laa 2})\subseteq Ess(f).$$

"$\Ra$"
Let  $f\in G_{2,k}^n$.  Let us set  $f=t\oplus g$, where
$$t=\bigoplus_{\alpha_1\alpha_2\ldots\alpha_n\in Eq_k^n}a_m.x_1^{\alpha_1} x_2^{\alpha_2}\ldots x_n^{\alpha_n}\quad\mbox{  and  }\quad g=\bigoplus_{\beta_1\beta_2\ldots\beta_n\notin Eq_k^n} a_r.x_1^{\beta_1} x_2^{\beta_2}\ldots  x_n^{\beta_n},$$
with  $m=\o{\alpha_1\alpha_2\ldots\alpha_n}$ and $r=\o{\beta_1\beta_2\ldots\beta_n}$.

Note that such representation of $f$ can be obtained after a suitable reordering of the conjunctions in its $SC$-form.

Clearly, $g_{i\laa j}=0$ for all $i,j\in \{1,\ldots,n\}$, $i\neq j$.

From Theorem \ref{t5.5} it follows that $f_{i\laa j}$ is  totally symmetric  and  $Ess(f_{i\laa j})=X_n\setminus\{x_i,x_j\}$   for all $i,j\in \{1,\ldots,n\}$,  $i\neq j$.   Since $f_{i\laa j}=t_{i\laa j}$ from Theorem \ref{t5.5} we can conclude that  $t_{i\laa j}$ is totally symmetric. We have to show that $t$ is totally symmetric, also. 
Since $Ess(t_{i\laa j})=X_n\setminus\{x_i,x_j\}$ there is an $n-2$-ary function $h$ such that $t_{i\laa j}=h$, where 
 $h$ is a totally symmetric function (according to Theorem  \ref{t5.5}) which depends essentially on all of its variables.  By $t(\alpha_1,\ldots,\alpha_n)=0$ when $\alpha_1\ldots\alpha_n\notin Eq_k^n$  it suffices to prove that 
\begin{eqnarray}\label{eq5.1}t(\alpha_1,\ldots,\alpha_{j-1},\beta,\alpha_{j+1},\ldots,\alpha_{i-1},\beta,\alpha_{i+1},\ldots,\alpha_n)=\nonumber \\
h(\alpha_1,\ldots,\alpha_{j-1},\alpha_{j+1},\ldots,\alpha_{i-1},\alpha_{i+1},\ldots,\alpha_n),\end{eqnarray}
for all $i,j\in \{1,\ldots,n\}$,  $i\neq j$.

First we shall prove (\ref{eq5.1}) for $j=1$ and $i=3$. Since $x_1\notin Ess(f_{3\laa 1})$ we obtain
$$t(\alpha_1,\alpha_2,\alpha_1,\ldots,\alpha_n)=t(\alpha_2,\alpha_2,\alpha_2,\ldots,\alpha_n)=h(\alpha_2,\ldots,\alpha_n)$$  as desired.

In a similar way we might show (\ref{eq5.1}) for $j=2$ and $i=3$.  As in the proof of Theorem \ref{t5.5} we may reorder the variables and  show  (\ref{eq5.1}) for all $i,j\in \{1,\ldots,n\}$,  $i\neq j$.

 \hfill ~~~\end{proof}
\begin{corollary} \label{c5.1}  Let $f\in G_{2,k}^n$, $3<n< k$ and $x_v\notin Ess(f_{u\laa v})$ for all $v,u\in \{1,\ldots,n\}$, $u\neq v$. Then there exists an $n-2$-ary totally symmetric function $h$ such that \\
\begin{tabular}{ll} $f_{i\laa j}$&$=f(x_1,\ldots,x_{j-1},x_j,x_{j+1},\ldots,x_{i-1},x_j,x_{i+1},\ldots,x_n)$\\ &$=
h(x_1,\ldots,x_{j-1},x_{j+1},\ldots,x_{i-1},x_{i+1},\ldots,x_n),$ \end{tabular}\\
for all $i,j\in \{1,\ldots,n\}$,  $i\neq j$.
 \end{corollary}
 \begin{remark}\label{rem5.1}
 We shall use some notation  and results from \cite{ber} and \cite{ros}.

 Let $Sub(K)$ denote the set of all subsets of $K$. Define $oddsupp: K^n\to Sub(K)$  as follows
 $$oddsupp(\alpha_1,\ldots,\alpha_n):=\{\alpha_v:  |\{u: \ 1\leq u\leq n\   \&\   \alpha_u=\alpha_v\}|\ \mbox{ is odd}\}.$$

A function $f$ is determined by $oddsupp$ if there exists a function $f^*: Sub(K)\to K$ such that $f=f^*\circ oddsupp$. It is easy to see that the functions which are determined by   $oddsupp$ are totally symmetric, too.
Then from Theorem 16 \cite{mig1} it follows that the function $t$ defined  in Theorem \ref{th5.6} is determined by  $oddsupp$.
\end{remark}

\section{The class $G_{2,k}^n$ with $2\leq n\leq 3\leq k$}\label{sec5}
~~\\

Let us note that the class $G_{2,k}^2$ is described in Section \ref{sec3} by Theorem \ref{t3.1}. Thus $f\in  G_{2,k}^2$ if and only if 
$$f=[\bigoplus_{\alpha\neq \beta}
a_{r} x_1^{\alpha} x_2^{\beta}]\oplus a_0.(x_1^0x_2^0\oplus\ldots\oplus x_1^{k-1}x_2^{k-1})$$
where at least two
among  the coefficients $\{a_0\}\cup\{a_{r}\ |\
r=\o{\alpha\beta},\  \& \ \alpha\neq\beta\}$
 are distinct.

\begin{example}~ \label{e3.2}
Let   $f\in P_3^3$ be the following   function
 $$f=x_1^0x_2^0x_3^0\oplus x_1^0x_2^0x_3^1\oplus x_1^0x_2^0x_3^2\oplus
  x_1^0x_2^1x_3^0\oplus x_1^0x_2^2x_3^0\oplus
 x_1^1x_2^0x_3^0\oplus x_1^2x_2^0x_3^0.$$
It is easy to check that
 $f_{2\laa 1}= f_{3\laa 1}=x_1^0$,  $f_{3\laa
2}=x_2^0$ and hence  $f\in G_{2,3}^3$.

Let  $g\in P_3^3$ be the following   function
$$g=x_1^0x_2^0x_3^0\oplus x_1^0x_2^1x_3^1\oplus x_1^0x_2^2x_3^2\oplus
x_1^1x_2^0x_3^1\oplus x_1^1x_2^1x_3^0\oplus
 x_1^2x_2^0x_3^2\oplus x_1^2x_2^2x_3^0.$$
 It is not difficult to see that  $g_{2\laa 1}=x_3^0$,
 $g_{3\laa 1}=x_2^0$, $g_{3\laa 2}=x_1^0$ and hence $g\in G_{2,3}^3$.

 Let $h\in P_3^3$ be the following   function
 $$h=x_1^0x_2^0\oplus x_1^0x_2^1x_3^1\oplus x_1^0x_2^2x_3^2\oplus
  x_1^1x_2^0x_3^1\oplus x_1^2x_2^0x_3^2.$$
It is clear that  
 $h_{2\laa 1}= h_{3\laa 2}=x_1^0$, $h_{3\laa
1}=x_2^0$ and hence  $h\in G_{2,3}^3$.

   Let $r\in P_3^3$ be the following   function
 $$r=x_2^0\oplus 2.x_1^1x_2^0x_3^2\oplus 2.x_1^2x_2^0x_3^1.$$
It is clear that  
 $r_{3\laa 1}= r_{3\laa 2}=x_2^0$, $r_{2\laa
1}=x_1^0$ and hence  $r\in G_{2,3}^3$.
  
  Clearly, the functions $f,g$ and $h$ do not satisfy Theorem \ref{th31.7} for $n=3$, and the functions  $f,h$, and $r$ do not satisfy Theorem \ref{th5.6} for $n=3$, but $g\in G_{2,3}^{3,-}$, and $r\in G_{2,3}^{3,+}$.

\end{example}

Example \ref{e3.2} shows that the case $(p,n)= (2,3)$  is a  special case in studying the functions with non-trivial arity gap.

 In this section we shall pay attention to  description of the class $G_{2,k}^3$ with $k>2$.

 %  \begin{example}~\label{e5.1}~

\begin{lemma}\label{l5.1}
If $f\in G_{2,k}^3$ then $ess(f_{i\laa j})=1$ for all $i,j\in\{1,2,3\}$, $i\neq j$.
\end{lemma}
\begin{proof}
Since  $f\in G_{2,k}^3$  there is an identification minor of $f$ which depends essentially on one variable and let us suppose
$ess(f_{2\laa 1})=1$. There are two possibilities.

$(i)$  $Ess(f_{2\laa 1})=\{x_3\}$. Then there are three constants $\alpha_1,\alpha_3,\beta_3\in K$ such that
$f(\alpha_1,\alpha_1,\alpha_3)\neq f(\alpha_1,\alpha_1,\beta_3)$.

$(ii)$  $Ess(f_{2\laa 1})=\{x_1\}$. Then there exist three constants $\mu_1,\mu_2,\nu_3\in K$, such that
$f(\mu_1,\mu_1,\nu_3)\neq f(\mu_2,\mu_2,\nu_3)$.

Now let us suppose the lemma is false and $ess(f_{3\laa 1})=0$. This implies that for all four constants $\delta_1,\varepsilon_1,\sigma_2,\psi_2\in K$ we have $f(\delta_1,\sigma_2,\delta_1)=f(\varepsilon_1,\psi_2,\varepsilon_1)$ which contradicts the both inequalities, listed above in $(i)$ and $(ii)$. The similar arguments work if we suppose  $ess(f_{3\laa 2})=0$.
 \hfill ~~~ \end{proof}

We are going to describe the properties of the functions from  $G_{2,3}^3$. For this  we need  the following auxiliary functions.
\begin{eqnarray}~~\label{eq23} s(x_1,x_2):=\bigoplus_{\beta=\alpha}x_1^\beta x_2^\alpha,\quad
u^{(\alpha)} (x_1,x_2):=\bigoplus_{\beta\neq\alpha}x_1^\beta x_2^\beta,\nonumber \\
v^{(\alpha)} (x_1,x_2):=\bigoplus_{\beta\neq\alpha}x_1^\alpha x_2^{\beta} ~~~~~~\end{eqnarray}
For example $u^{(1)}(x_3,x_2)=x_3^0x_2^0\oplus x_2^2x_2^2$ and $v^{(1)}(x_3,x_2)=x_3^1x_2^0\oplus x_3^1x_2^2$.

The following representation of the elements in $G_{2,3}^3$ should be also compared with Theorem 17(iii) in \cite{mig1}.
\begin{theorem}\label{t5.1}
If $f\in G_{2,3}^3,$ 
then $f$ has to  be  represented in one of the following special forms, up to permutation of variables:
\begin{equation}\label{eq24}
 f=\bigoplus_{i=0}^2 a_0^{(i)}[x_3^i.s(x_1,x_2)
 \oplus  x_2^i.u^{(i)}(x_1,x_3)
 \oplus x_1^i.u^{(i)}(x_2,x_3)]\oplus p_3(x_1,x_2,x_3),
\end{equation}

\begin{equation}~\label{eq21}
f=\bigoplus_{i=0}^2 a_0^{(i)}[x_1^i.x_2^i
 \oplus  x_1^i.u^{(i)}(x_2,x_3)
 \oplus x_2^i.u^{(i)}(x_1,x_3)]\oplus p_3(x_1,x_2,x_3),
\end{equation}

\begin{equation}\label{eq22}
 f=\bigoplus_{i=0}^2 a_0^{(i)}[x_1^i.x_2^i
 \oplus  x_2^i.v^{(i)}(x_3,x_1)
 \oplus x_2^i.u^{(i)}(x_1,x_3)]\oplus p_3(x_1,x_2,x_3),\end{equation}

\begin{equation}\label{eq25}
 f=\bigoplus_{i=0}^2 a_0^{(i)}[x_1^i.x_2^i
 \oplus  x_1^i.v^{(i)}(x_3,x_2)
 \oplus x_2^i.v^{(i)}(x_3,x_1)]\oplus p_3(x_1,x_2,x_3),
\end{equation}
such that at least two among the coefficients $a_0^{(0)}, a_0^{(1)}, a_0^{(2)}$  are different and  $p_3$ are arbitrary functions defined by (\ref{eq3.3}).
\end{theorem}
\begin{proof}
According to Lemma \ref{l5.1}, for $f\in G_{2,3}^3$ we need $ess(f_{2\laa 1})=ess(f_{3\laa 1})=ess(f_{3\laa 2)}=1$. This is equivalent to the following conjunction of three disjunctions:
$$ \big( x_1\notin Ess(f_{2\laa 1})\mbox{   or    } x_3\notin Ess(f_{2\laa 1})\big) \  \& \
 \big( x_1\notin Ess(f_{3\laa 1})\mbox{   or    } x_2\notin Ess(f_{3\laa 1})\big)$$ $$ \&\
 \big( x_2\notin Ess(f_{3\laa 2})\mbox{   or    } x_1\notin Ess(f_{3\laa 2})\big).$$ Thus we obtain the following linear systems of equations for the coefficients of the functions from $G_{2,3}^3$.
\begin{eqnarray}~\label{eq26}
(a)~~~& (a_0^{(0)},a_0^{(1)},a_0^{(2)})=(a_4^{(0)},a_4^{(1)},a_4^{(2)})=(a_8^{(0)},a_8^{(1)},a_8^{(2)})~~~~~~\nonumber \\
(b)~~~& (a_0^{(0)},a_4^{(0)},a_8^{(0)})=(a_0^{(1)},a_4^{(1)},a_8^{(1)})=(a_0^{(2)},a_4^{(2)},a_8^{(2)}),\quad\quad
\end{eqnarray}

\begin{eqnarray}~\label{eq27}
(a)~~~& (a_0^{(0)},a_1^{(0)},a_2^{(0)})=(a_3^{(1)},a_4^{(1)},a_5^{(1)})=(a_6^{(2)},a_7^{(2)},a_8^{(2)})~~~~~~\nonumber \\
(b)~~~&  (a_0^{(0)},a_3^{(1)},a_6^{(2)})=(a_1^{(0)},a_4^{(1)},a_7^{(2)})=(a_2^{(0)},a_5^{(1)},a_8^{(2)}),\quad\quad
\end{eqnarray}

\begin{eqnarray}~\label{eq28}
(a)~~~&  (a_0^{(0)},a_3^{(0)},a_6^{(0)})=(a_1^{(1)},a_4^{(1)},a_7^{(1)})=(a_2^{(2)},a_5^{(2)},a_8^{(2)})~~~~~~\nonumber \\
(b)~~~& (a_0^{(0)},a_1^{(1)},a_2^{(2)})=(a_3^{(0)},a_4^{(1)},a_5^{(2)})=(a_6^{(0)},a_7^{(1)},a_8^{(2)}).\quad\quad
\end{eqnarray}

Now any solution is determined by a word (string) over the alphabetic $\{a,b\}$ with length 3. For example let us consider the string $aba$. Then we use equations $a$ from (\ref{eq26}),   $b$ from (\ref{eq27}) and  $a$ from (\ref{eq28}).

Then  we obtain
\begin{eqnarray*}
a_{0}^{(0)}=a_{4}^{(0)}=a_{8}^{(0)}=a_{1}^{(0)}=a_{2}^{(0)}=a_{1}^{(1)}=a_{2}^{(2)}\\
a_{0}^{(1)}=a_{4}^{(1)}=a_{8}^{(1)}=a_{3}^{(1)}=a_{5}^{(1)}=a_{3}^{(0)}=a_{5}^{(2)}\\
a_{0}^{(2)}=a_{4}^{(2)}=a_{8}^{(2)}=a_{6}^{(2)}=a_{7}^{(2)}=a_{6}^{(0)}=a_{1}^{(1)}.
\end{eqnarray*}
Hence
$$ f=\bigoplus_{i=0}^2 a_0^{(i)}[x_1^i.x_3^i
 \oplus  x_1^i.u^{(i)}(x_2,x_3)
 \oplus x_3^i.u^{(i)}(x_1,x_2)],$$ i.e. $f$ is as the function presented by (\ref{eq21}) where the variables $x_2$ and $x_3$ are permuted.

  In this way we might generate $2^3=8$ linear systems for the coefficients.

 The remaining seven cases of strings, might be checked in the same way.
 We will summarize the results. The functions in the form (\ref{eq24}) are generated by the string $aaa$; (\ref{eq21}) -  by  $aba$, $baa$ and $aab$;  (\ref{eq22}) - by  $abb$, $bba$ and $bab$ and (\ref{eq25}) - by  $bbb$.

\hfill ~~~ \end{proof}

 \begin{proposition}~\label{p1}
 $|G_{2,3}^3|=139968.$
 \end{proposition}
\begin{proof}
 Clearly, each of the functions in equations  (\ref{eq24}) - (\ref{eq25}) can be written as follows
 $$ f=\big[\bigoplus_{i=0}^2 a_0^{(i)}.g^{(i)}(x_1,x_2,x_3)\big]\oplus p_3(x_1,x_2,x_3),$$
 where $p_3$ are the functions defined in (\ref{eq3.3}) when $k=3$. It is easy to show that the number of all such functions is equal to  $3^{3!}=729$.
 
The functions
 $g^{(i)}$ from the equations (\ref{eq24}) and (\ref{eq25}) are totally symmetric, but these from (\ref{eq21}) and (\ref{eq22}) are symmetric with respect to two of their arguments, only. Thus if the coefficients $a_0^{(i)}$ are fixed then there exist one function $$\bigoplus_{i=0}^2 a_0^{(i)}.g^{(i)}(x_1,x_2,x_3)$$ determined by  (\ref{eq24}), one - by (\ref{eq25}), three functions are determined by (\ref{eq21}) and three - by (\ref{eq22}). So, there are 8 such functions. The triples $(a_0^{(0)},a_0^{(1)},a_0^{(2)})$ can be chosen in $3^3-3=24$ ways (the triples $(0,0,0)$, $(1,1,1)$ and $(2,2,2)$ are forbidden). Hence there are $8.24=192$ functions $\bigoplus_{i=0}^2 a_0^{(i)}.g^{(i)}(x_1,x_2,x_3)$ which satisfied (\ref{eq24}) - (\ref{eq25}). Hence  $|G_{2,3}^3|=192.729=139968.$
\hfill ~~~ \end{proof}

\begin{proposition}~\label{p5.2}
 $|G_{2,k}^3|=8.729.{k\choose 3}.(k^k-k)=5832.{k\choose 3}.(k^k-k).$
 \end{proposition}
 \begin{proof}
 Without any difficulties, excluding the more complex calculations, we might  generalize  results from  $G_{2,3}^3$ to $G_{2,k}^3$ for arbitrary $k$, $k\geq 3$. In this  case we obtain the same conjunction of three disjunctions to determine the functions from $G_{2,k}^3$. The difference is that in the equations (\ref{eq26})-(\ref{eq28}) participate $k$ tuples of $k$ coefficients and a function belongs to $G_{2,k}^3$ if and only if it can be represented as in (\ref{eq24})-(\ref{eq25}) as the sums are extended up to $k-1$ instead of $2$ in the case $G_{2,3}^3$. All other arguments work,  here also.

\hfill ~~~ \end{proof}

\section*{  Acknowledgements}

We are  grateful to our  colleagues K. Chimev, K. Denecke and D. Kovachev for useful discussions and their comments.


\begin{thebibliography} {20}
\bibitem{ber} J. Berman and A. Kisielewicz, {On the number of operations in a clone,}
Proc. Amer. Math Soc., 122 (1994), pp. 359-369.
\bibitem{bre} Yu. Breitbart, {On the essential variables of
  functions in the algebra of logic,}
 Dokl. Acad. Sci. USSR, 172, vol. 1, (1967), pp. 9-10 (in Russian).
\bibitem{ch51} K. Chimev, { Separable sets of arguments of functions},
MTA SzTAKI Tanulmanyok, 180, (1986), 173 p.
\bibitem{ch3} {\rm K. Chimev,  { On some properties of functions, }
Colloquia Mathematica Societatis Janos Bolyai, Szeged, (1981),
pp. 97-110.}
\bibitem{mig0} M. Couceiro, E. Lehtonen, On the arity gap of finite functions: results and applications,   Int. Conf. on Relations, Orders and Graphs: Interaction with Computer Science, Nouha Editions, Sfax, (2008), pp. 65-72, (http://www.math.tut.fi/algebra/papers/ROGICS08-CL.pdf).
\bibitem{mig1} M. Couceiro, E. Lehtonen, Generalizations of Swierczkowski's lemma and the arity gap of finite functions, Discrete Mathematics, (2009), doi:10.1016/j.disc.2009.04.009.

\bibitem{den} K. Denecke, J. Koppitz, Essential variables in hypersubstitutions,  Algebra Universalis, 46 (2001), pp. 443-454

\bibitem{kov} D. Kovachev, { On a class of discrete functions},  Acta Cybernetica, vol. 17, No. 3, Szeged, (2006), pp. 513-519.
\bibitem{lup} O. Lupanov,
  { On a class of schemes of functional elements,}
   Problemi Kybernetiki, 9, (1963), pp. 333-335 (in Russian).
\bibitem{sal}{\rm A. Salomaa,  { On essential variables of functions,
especially in the algebra of logic, } Annales Academia Scientiarum
Fennicae, Ser. A, 333, (1963), pp. 1-11.}
\bibitem{sh51}  Sl. Shtrakov, K. Denecke, { Essential variables and
separable sets in universal algebra}, Taylor \& Francis,
Multiple-Valued Logic, vol. 8, No. 2, (2002),
pp. 165-182.
\bibitem{sh52}  Sl. Shtrakov,
Essential variables and positions in terms,  Algebra Universalis, vol. 61, No. 3-4  (2009),  pp. 381-397.
\bibitem{s1}
Sl. Shtrakov, { Tree automata and essential input variables},
Contributions to General Algebra 13, Verlag Johannes Heyn,
Klagenfurt, (2001), pp.309-320.
\bibitem{s2} Sl. Shtrakov, { Essential arity gap of Boolean
functions},{ Serdica Journal of Computing,} vol.2, No. 3, (2008), pp. 249-266.
\bibitem{ros} R. Willard, { Essential arities of term operations in finite algebras},
Discrete Mathematics, 149 (1996), pp. 239-259.

\end{thebibliography}
\end{document}